Comment on the paper "Magnetohydrodynamic non-Darcy mixed convection heat transfer from a vertical heated plate embedded in a porous medium with variable porosity, by Dulal Pal, Commun Nonlinear Sci Numer Simulat, 15 (2010) 3974-3987 "

Asterios Pantokratoras
School of Engineering, Democritus University of Thrace,
67100 Xanthi – Greece
e-mail:apantokr@civil.duth.gr

In the above paper the author treats the boundary layer flow along a vertical flat plate, immersed in a Darcy-Brinkman –Forchheimer porous medium. The porosity and the permeability of the porous medium are variable across the boundary layer. In addition a magnetic field with constant strength is applied normal to the plate. The fluid temperature at the plate is constant and different from that of the ambient fluid. This temperature difference creates a buoyancy force and the flow is characterized as mixed convection. The partial differential equations of the boundary layer flow ( Eqs. 4-6 in his paper) are transformed into ordinary differential equations and subsequently are solved with the Runge-Kutta Fehlberg method. The results are presented in two tables and 11 figures.

The partial differential equations of this problem, using the same symbols by Pal (2010) are

$$\frac{\partial u}{\partial x} + \frac{\partial v}{\partial y} = 0 \qquad (1)$$

$$u\frac{\partial u}{\partial x} + v\frac{\partial u}{\partial y} = g\beta(T-T_\infty) + \frac{\mu_{ef}}{\rho}\frac{\partial^2 u}{\partial y^2} + \frac{\mu\varepsilon(y)}{\rho k(y)}(U_0 - u) + \frac{C_b \varepsilon^2(y)}{\sqrt{k(y)}}(U_0^2 - u^2) + \frac{\sigma_m B_0^2}{\rho}(U_0 - u) \qquad (2)$$

$$u\frac{\partial T}{\partial x}+v\frac{\partial T}{\partial y}=\frac{\partial}{\partial y}\left(a(y)\frac{\partial T}{\partial y}\right)+u\left[\frac{\mu\varepsilon(y)}{\rho c_p k(y)}(u-U_0)+\frac{C_b\varepsilon^2(y)}{c_p\sqrt{k(y)}}(u^2-U_0^2)-\frac{\mu_{ef}}{\rho c_p}\frac{\partial^2 u}{\partial y^2}\right] \quad (3)$$

subject to boundary conditions

$$u=0, v=0, T=T_w \text{ at y=0} \quad (4)$$

$$u \to U_0, T \to T_\infty \quad \text{as} \quad y \to \infty \quad (5)$$

where u and v are the velocity components in the x and y- directions, g is the gravitational acceleration, β is the thermal expansion coefficient, T is the fluid temperature, $\mu_{ef}$ is the fluid effective viscosity, μ is the normal viscosity, ρ is the fluid density, ε(y) is the medium porosity, k(y) is the permeability of the porous medium, $C_b$ is the Forchheimer coefficient, $\sigma_m$ is the electrical conductivity, $B_0$ is the strength of the magnetic field, a(y) is the thermal diffusivity and $c_p$ is the specific heat at constant pressure.

The transformed momentum equation given by Pal (2010, his Eq. 17) is as follows :

$$f'''+\frac{1}{2}ff''+\frac{Gr}{Re^2}\theta+\frac{a^*}{\sigma Re}\frac{1+d^*e^{-\eta}}{1+de^{-\eta}}(1-f')+M^2(1-f')+\frac{\beta^*(1+d^*e^{-\eta})}{(1+de^{-\eta})^{1/2}}(1-f'^2)=0 \quad (6)$$

where

$$Gr=\frac{g\beta(T_w-T_\infty)x^3}{\upsilon^2} \quad (7)$$

$$Re=\frac{U_0 x}{\upsilon} \quad (8)$$

$$\theta=\frac{T-T_\infty}{T_w-T_\infty} \quad (9)$$

$$a^*=\frac{\mu}{\mu_{ef}} \quad (10)$$

$$\sigma = \frac{k_0}{x^2 \varepsilon_0} \tag{11}$$

$$M^2 = \frac{\sigma_m B_0^2 x}{\rho U_0} \tag{12}$$

$$\beta^* = \frac{C_b \varepsilon_0^2 x}{k_0^{1/2}} \tag{13}$$

$$\eta = \frac{y}{x} \mathrm{Re}^{\frac{1}{2}} \tag{14}$$

$$k(\eta) = k_0 (1 + d e^{-\eta}) \tag{15}$$

$$\varepsilon(\eta) = \varepsilon_0 (1 + d^* e^{-\eta}) \tag{16}$$

In Eq. (6) prime signifies differentiation with respect to $\eta$. The quantity Gr is the Grashof number, Re is the Reynolds number, $\theta$ is the dimensionless temperature, $\alpha^*$ is the ratio of viscosities, $\sigma$ is a Darcy parameter, the quantity $M^2$ is the magnetic parameter, $\beta^*$ is the Forchheimer parameter and $\eta$ is the transverse similarity variable. The variation of permeability and porosity across the boundary layer is given by Eqs. (15) and (16) where d and $d^*$ are constants. All the above quantities have been taken from the work of Pal (2010) with the same notation.

Let us consider the flow in a clear fluid, without porous medium, and without magnetic field. In this case the quantities $M^2$ and $\beta^*$ are zero and the quantity $\sigma$ is infinite. Therefore the momentum Eq. (6) takes the form

$$f''' + \frac{1}{2} f f'' + \frac{Gr}{\mathrm{Re}^2} \theta = 0 \tag{17}$$

which must express the mixed convection flow along a vertical plate placed inside a constant free stream in a clear fluid. This problem has been solved by Ramachandran et al. (1985) both numerically and experimentally with a very good agreement between predictions and measurements. However, the corresponding momentum equation for this problem given by Ramachandran et al. (1985, Eq. 7) is

4$$F'''+\frac{1}{2}FF''+\xi\theta = \xi(F'\frac{\partial F'}{\partial \xi} - F''\frac{\partial F}{\partial \xi}) \tag{18}$$

where $\xi = \frac{Gr}{Re^2}$. The quantities $f$ and $\eta$ in Pal (2010) are identical with the quantities $F$ and $\eta$ in the work of Ramachandran et al. Comparing the momentum Eqs. (17) and (18) it is clear that they are quite different. The two terms in the right hand side of Eq. (18) are missing from Eq. (17) which apparently is wrong. Taking into account this fact the credibility of the entire results presented by Pal (2010) is doubtful.